\documentclass[article]{aa}
\usepackage{graphicx}
\usepackage[varg]{txfonts}
\begin{document}

\title{Deriving physical parameters of unresolved star clusters}
\subtitle{IV. The M33 star cluster system}
\author{P. de Meulenaer\inst{1,2} \and D. Narbutis\inst{1,2} \and T. Mineikis\inst{1,2} \and V. Vansevi\v{c}ius\inst{1,2}}

\institute{Vilnius University Observatory, \v{C}iurlionio 29, Vilnius LT-03100, Lithuania\\
\email{philippe.demeulenaer@ff.stud.vu.lt, vladas.vansevicius@ff.vu.lt} \and
Center for Physical Sciences and Technology, Savanori\c{u} 231, Vilnius LT-02300, Lithuania}

\date{Submitted 18 May 2015 / Accepted 29 June 2015}

%%%%%%%%%%%%%%%%%%%%%%%%%%%%%%%%%%%%%%%%%%%%%%%%%%%%%%%%%
\abstract
%%%%%%%%%%%%%%%%%%%%%%%%%%%%%%%%%%%%%%%%%%%%%%%%%%%%%%%%%
{When trying to derive the star cluster physical parameters of the M33 galaxy using broad-band unresolved ground-based photometry, previous studies mainly made use of simple stellar population models, shown in the recent years to be oversimplified.}
{In this study, we aim to derive the star cluster physical parameters (age, mass, and extinction; metallicity is assumed to be LMC-like for clusters with age below 1\,Gyr and left free for older clusters) of this galaxy using models that take stochastic dispersion of cluster integrated colors into account.}
{We use three recently published M33 catalogs of cluster optical broad-band photometry in standard $UBVRI$ and in CFHT/MegaCam $u^{*}g'r'i'z'$ photometric systems. We also use near-infrared $JHK$ photometry that we derive from deep 2MASS images. We derive the cluster parameters using a method that takes into account the stochasticity problem, presented in previous papers of this series.}
{The derived differential age distribution of the M33 cluster population is composed of a two-slope profile indicating that the number of clusters decreases when age gets older. The first slope is interpreted as the evolutionary fading phase of the cluster magnitudes, and the second slope as the cluster disruption. The threshold between these two phases occurs at $\sim$300\,Myrs, comparable to what is observed in the M31 galaxy. We also model by use of artificial clusters the ability of the cluster physical parameter derivation method to correctly derive the two-slope profile for different photometric systems tested.}
{}


\maketitle

%%%%%%%%%%%%%%%%%%%%%%%%%%%%%%%%%%%%%%%%%%%%%%%%%%%%%%%%%
\section{Introduction}
%%%%%%%%%%%%%%%%%%%%%%%%%%%%%%%%%%%%%%%%%%%%%%%%%%%%%%%%%
There is a current need for an accurate catalog for the star cluster system of the Triangulum galaxy, or Messier 33 (M33), as it could be used as a constraint for the derivation of the galaxy star formation history. Several other reasons encourage the study of this particular star cluster system. The nearly face-on inclination \citep[$i = 56$ degrees,][]{Regan1994} of M33 reduces extinction effects for the majority of its cluster population, situated in the disk. Also, M33 is the only close late-type spiral galaxy, situated at a distance of 867 kpc \citep[][distance modulus of $(m-M)_{0} = 24.69$]{Galleti2004}, making its star cluster system accessible to ground-based telescopes for integrated photometric and spectroscopic studies and to the Hubble Space Telescope (HST) for resolved measurements. While other star cluster systems of Local Group galaxies have received considerable attention, as in the case of M31 and the Magellanic Clouds, the M33 star cluster system has not been studied as much. Therefore an extended knowledge of its star cluster system would improve the understanding of the relationship between star clusters and their host galaxies. 

The M33 star cluster system has nevertheless been studied for a long time. Several teams \citep{Hiltner1960,Kron1960,Melnick1978,Christian1982,Christian1988,Mochejska1998} contributed to the building of a catalog of star clusters in M33 using ground-based unresolved photometry in optical passbands. \cite{Chandar1999,Chandar1999a,Chandar1999b,Chandar2001} used the WFPC2 camera onboard HST to detect 102 additional clusters. They derived their physical parameters using integrated photometry that were compared with simple stellar population (SSP) models, which are ideal star cluster models in the sense that they are based on a continuously sampled initial mass function (IMF). \cite{Sarajedini1998,Sarajedini2000,Sarajedini2007b} also used WFPC2 and ACS images, but derived the parameters using resolved color-magnitude diagrams for the most massive clusters. All studies before 2007 have been combined in a merged catalog by \cite{Sarajedini2007}. 

More recently, studies in the literature continue to use the SSP models to derive the M33 cluster physical parameters (age, mass, extinction, and metallicity). However several recent studies \citep[e.g., ][]{Santos1997, Deveikis2008,Fouesneau2010} have brought attention to the fact that these models with continuously sampled IMF, which are unphysical, are oversimplified and biased: 
\begin{itemize}
\item They are oversimplified because they do not take the natural dispersion of star cluster integrated colors into account, the so-called \emph{stochasticity problem}. In reality, the masses of stars are stochastically sampled following the stellar IMF, and this could be seen as a probability distribution function \citep{Santos1997}. Hence, two clusters with same physical parameters could host a different number of massive bright stars, resulting in very different integrated colors, especially in the case of young and low-mass clusters that contain only a few massive bright stars.
\item SSP models are biased because they do not even match the average of integrated color distributions for star clusters with mass $\log_{10}(M/M_{\odot}) \lesssim 4$ \citep[see, e.g., ][]{Fouesneau2010,Popescu2010,Silva_Villa2011}.
\end{itemize}
As a consequence, the derivation of star cluster parameters using the SSP models can lead to severe biases, as was demonstrated by \cite{Fouesneau2010} (see their Fig. 3) by use of artificial tests.

New models \citep[][hereafter Papers I, II, and III]{Fouesneau2010,Popescu2010,daSilva2012,de_Meulenaer2013,de_Meulenaer2014,de_Meulenaer2015} take into account this natural dispersion of integrated colors due to the star mass stochastic sampling and allow the cluster parameters to be derived in a more realistic way, avoiding the strong biases of the SSP method.  

In this paper, we derive the physical parameters of a merged catalog of M33 star clusters, the first time ever using a stochastic method on these galaxy star clusters. To this end, we use three catalogs of clusters recently published in the literature: the \cite{SanRoman2010} catalog in the $u^{*}g'r'i'z'$ photometric system, and the \cite{Ma2012,Ma2013} and \cite{Fan2014} catalogs in the standard $UBVRI$ photometric system. We also use near-infrared photometry that we derive from deep 2MASS images.

The structure of the paper is the following. Section \ref{sec:cluster_data} presents the catalogs of clusters used in this study as well as the 2MASS photometry derivation. Section \ref{sec:parameters_determination} presents the derivation of the star cluster physical parameters using our developed method which takes stochastic dispersion of star cluster integrated colors into account. Section \ref{sec:artificial_test} presents the artificial tests used to estimate the reliability of our method for deriving the star cluster physical parameters and the global evolutionary fading and disruption timescales. The derivation of the physical parameters of the M33 cluster population and the derivation of the fading and disruption timescales are performed in Section \ref{sec:derived_parameters}. Conclusions are presented in Section \ref{sec:conclusions}.

%%%%%%%%%%%%%%%%%%%%%%%%%%%%%%%%%%%%%%%%%%%%%%%%%%%%%%%%%    
\section{The cluster data}
\label{sec:cluster_data}
%%%%%%%%%%%%%%%%%%%%%%%%%%%%%%%%%%%%%%%%%%%%%%%%%%%%%%%%%
Recently \cite{SanRoman2010} observed 803 M33 star clusters (599 candidates and 204 confirmed using the HST) using the 3.6 m Canada-France-Hawaii-Telescope (CFHT) and published a catalog in the MegaCam camera $u^{*}g'r'i'z'$ photometric system. Although their catalog also contained the cluster photometry converted to the $ugriz$ photometric system of the Sloan Digital Sky Survey (SDSS), we considered here the native MegaCam photometric system to avoid likely conversion approximation. 

Using archival images of the Local Group Galaxies Survey \cite[LGGS,][]{Massey2006} obtained using the 4 m Kitt Peak National Observatory telescope, \cite{Ma2012} built a catalog of 392 clusters, and \cite{Ma2013} added 234 others, all in the standard $UBVRI$ photometric system. \cite{Fan2014} also reanalyzed the LGGS photometry to publish a catalog of 708 clusters. In this paper, we adopt for the $UBVRI$ photometric system the \cite{Ma2012,Ma2013} photometry when available, and the \cite{Fan2014} photometry for the other clusters, correcting the \cite{Fan2014} photometric zero-points to the \cite{Ma2012,Ma2013} ones in order to have an homogeneous catalog. The zero-point correction coefficients for \cite{Fan2014} minus \cite{Ma2012,Ma2013} photometry are $\Delta V=-0.099$\,mag, $\Delta (U-V)=-0.091$\,mag, $\Delta (B-V)=-0.066$\,mag, $\Delta (V-R)=0.036$\,mag, and $\Delta (V-I)=0.086$\,mag, computed for respectively 289, 208, 208, 250, and 178 clusters with available photometry in common between \cite{Fan2014} and \cite{Ma2012,Ma2013} catalogs.

\begin{figure}
\centering
\includegraphics[scale=0.45]{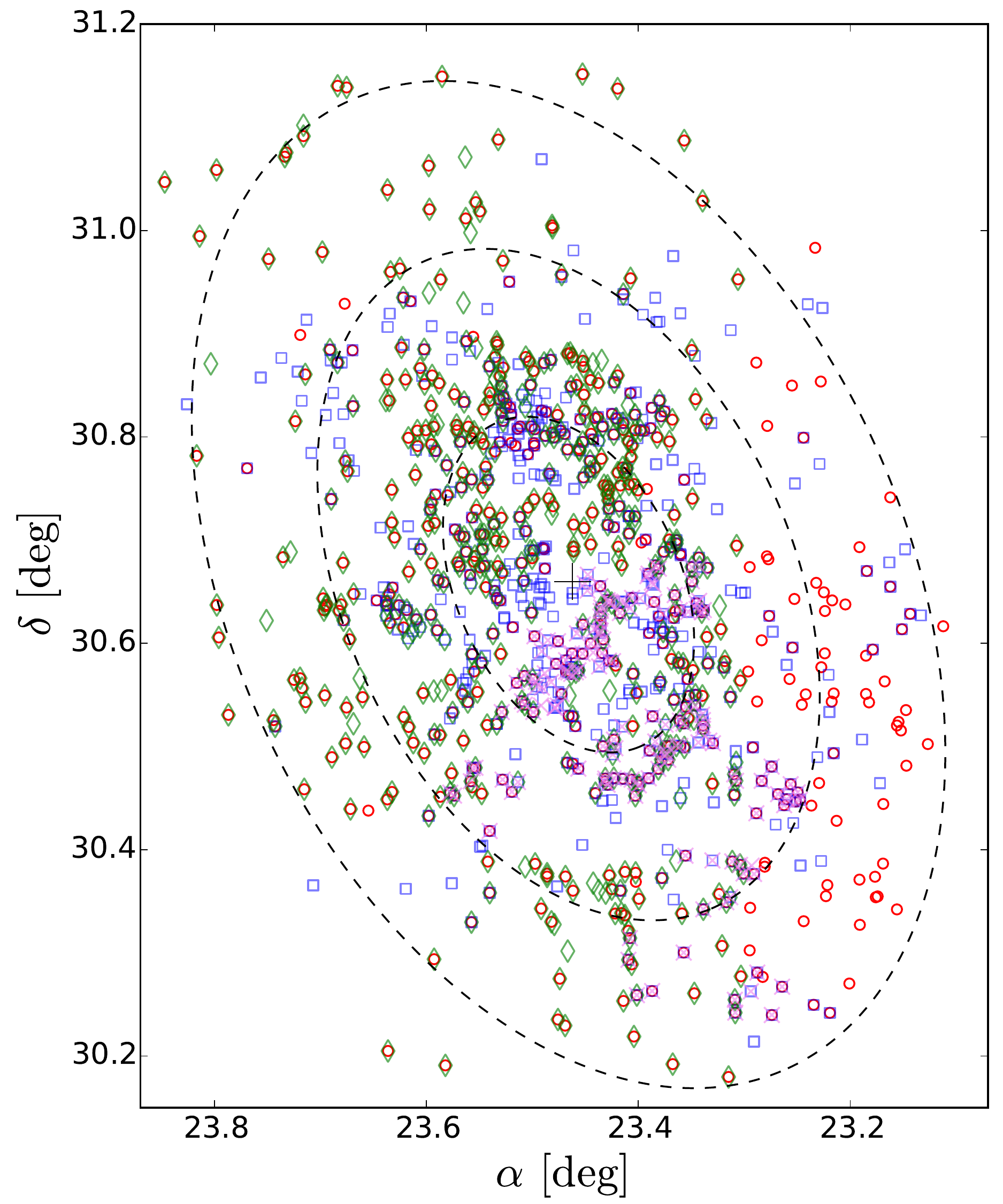}
\caption{\small The different catalogs of clusters used in this paper: \cite{Fan2014} in open red circles, \cite{Ma2012,Ma2013} in open blue squares, \cite{SanRoman2010} in green diamonds, \cite{SanRoman2009} in violet crosses. The three dashed ellipses have semi-major axes of 10\arcmin, 20\arcmin, and 30\arcmin\,to the center (marked as a large black plus symbol) and can be seen as circles of the same radii projected on the M33 disk.}
\label{fig:M33_positions_of_clusters}
\end{figure}

These catalogs include all objects published in the catalog of star clusters of \cite{Sarajedini2007}, who merged all the M33 star cluster catalogs published before 2007. The association of these catalogs in this paper results in a merged catalog of 910 objects, which is shown in Fig.\,\ref{fig:M33_positions_of_clusters} color-coded to show to which original catalog they belong.

\begin{figure*}
\centering
\includegraphics[scale=0.45]{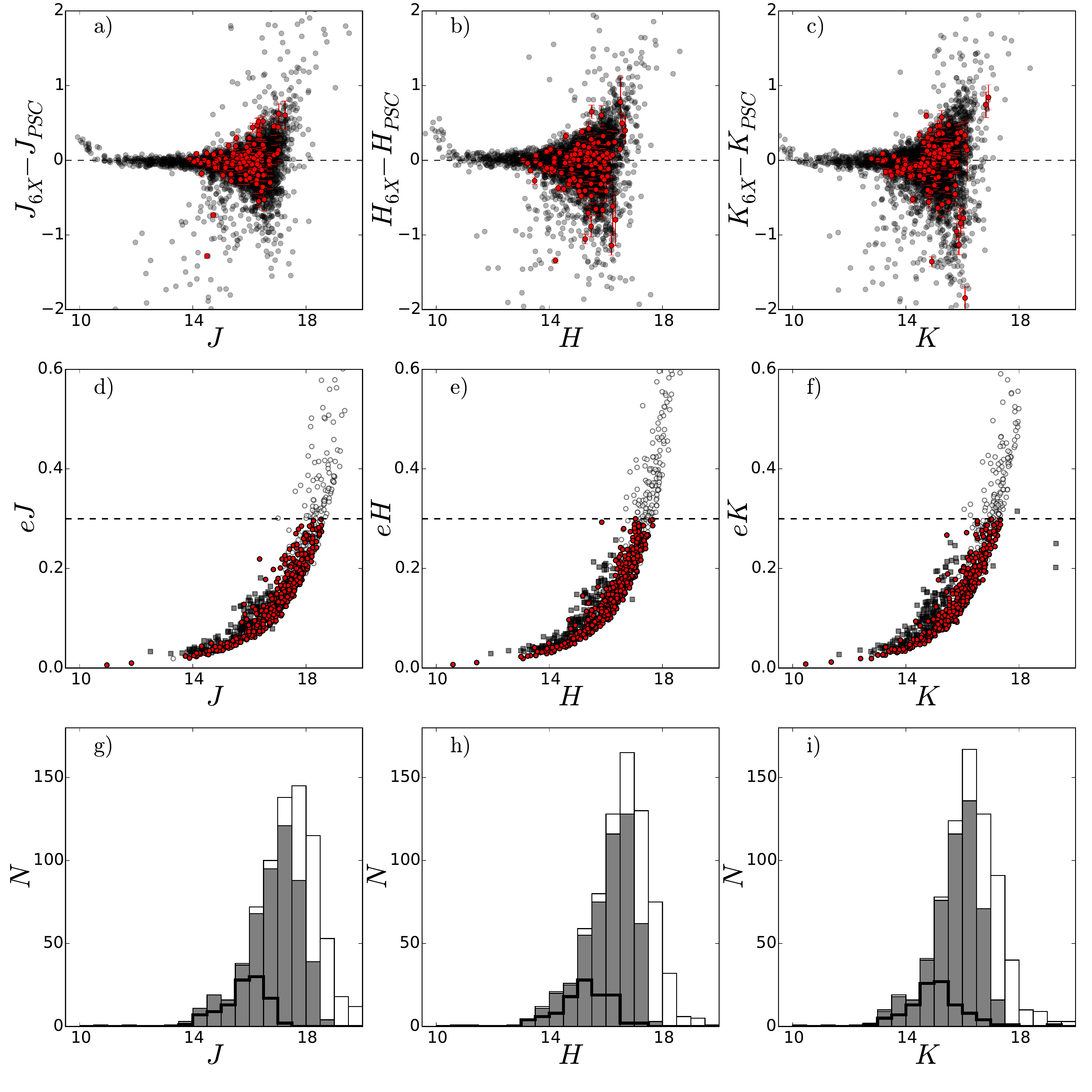} 
\caption{\small Top row: black circles show the comparison of the aperture photometry of stars derived in this work versus the PSC values, and red circles are the comparison of our aperture photometry of the star clusters versus the PSC values for the 109 clusters for which the PSC has aperture photometry. Central row: black squares show the photometric uncertainties provided by the PSC for 109 clusters and all circles show the photometric uncertainties derived for our aperture photometry for 758 clusters. The red circles are 502 clusters with uncertainties lower than the limit of 0.3\,mag (dashed line) in all $JHK$ passbands, and the open circles are the clusters for which the photometry does not satisfy this criteria. Bottom row: distributions of cluster brightness: white histograms contain the 758 clusters for which we derived photometry and the gray ones contain the 502 clusters that satisfy the photometric uncertainty criteria. The thick open histograms represent the 109 clusters contained in the PSC, for comparison. In each row, the situation is shown for $J$ (left panels), $H$ (central panels), and $K$ (right panels) passbands.}
\label{fig:2MASS_histograms_and_errors}
\end{figure*}

We supplemented optical data with near-infrared data by using deep Two Micron All Sky Survey (2MASS) $JHK$ images\footnote{Kindly made available by T.\,H.\,Jarrett, IPAC/Caltech} with exposure times 6 times longer (2MASS 6X) than the standard 2MASS ones. This results in a photometry approximatively 1.5\,mag deeper in 2MASS 6X than photometry derived in standard 2MASS images. The photometry of clusters was derived by use of aperture photometry using the standard IRAF/digiphot/apphot package with an aperture radius in the range 2\arcsec\,to 4\arcsec, and an aperture correction built using the curves of growth of a dozen relatively isolated clusters. By this process, we derived the $JHK$ photometry of 758 clusters. To ensure that the 2MASS 6X images were correctly calibrated, we also derived aperture photometry of stars, selected in the same region where clusters are located, and compared this derived aperture photometry to the stellar aperture photometry provided by the 2MASS Point Source Catalog (2MASS PSC)\footnote{http://irsa.ipac.caltech.edu}, which has been compiled using the standard 2MASS images. The first row of panels in Fig.\,\ref{fig:2MASS_histograms_and_errors} presents the comparison of aperture photometry of stars derived in this work versus the aperture photometry provided by the PSC (black circles) for the $JHK$ passbands. For most of stars the agreement is satisfactory. We noted that 109 clusters are included in the PSC and we could compare the aperture photometry that PSC provides for them with our aperture photometry, in red in the figure. The error bars reflect the photometric uncertainties derived in our aperture photometry. The photometric uncertainties of the 758 clusters derived in this work are shown in the central row of Fig.\,\ref{fig:2MASS_histograms_and_errors} (all circles), compared with the uncertainties of the 109 clusters for which PSC also provides aperture photometry (black squares). In this paper we use only the clusters with photometric uncertainties lower than 0.3\,mag in all $JHK$ passbands, indicated in the central row of panels of Fig.\,\ref{fig:2MASS_histograms_and_errors} by dashed lines. This selection results in 502 clusters with full $JHK$ photometry. The bottom row in Fig.\,\ref{fig:2MASS_histograms_and_errors} presents the distribution of the clusters in each of the $JHK$ passbands of our derived photometry for the whole sample of 758 clusters (in white) and for the sample of 502 clusters with uncertainties lower than 0.3\,mag in all $JHK$ passbands (in gray), compared to the distribution of 109 clusters that are included in the PSC (open thick histogram).

\begin{figure*}
\includegraphics[scale=0.7]{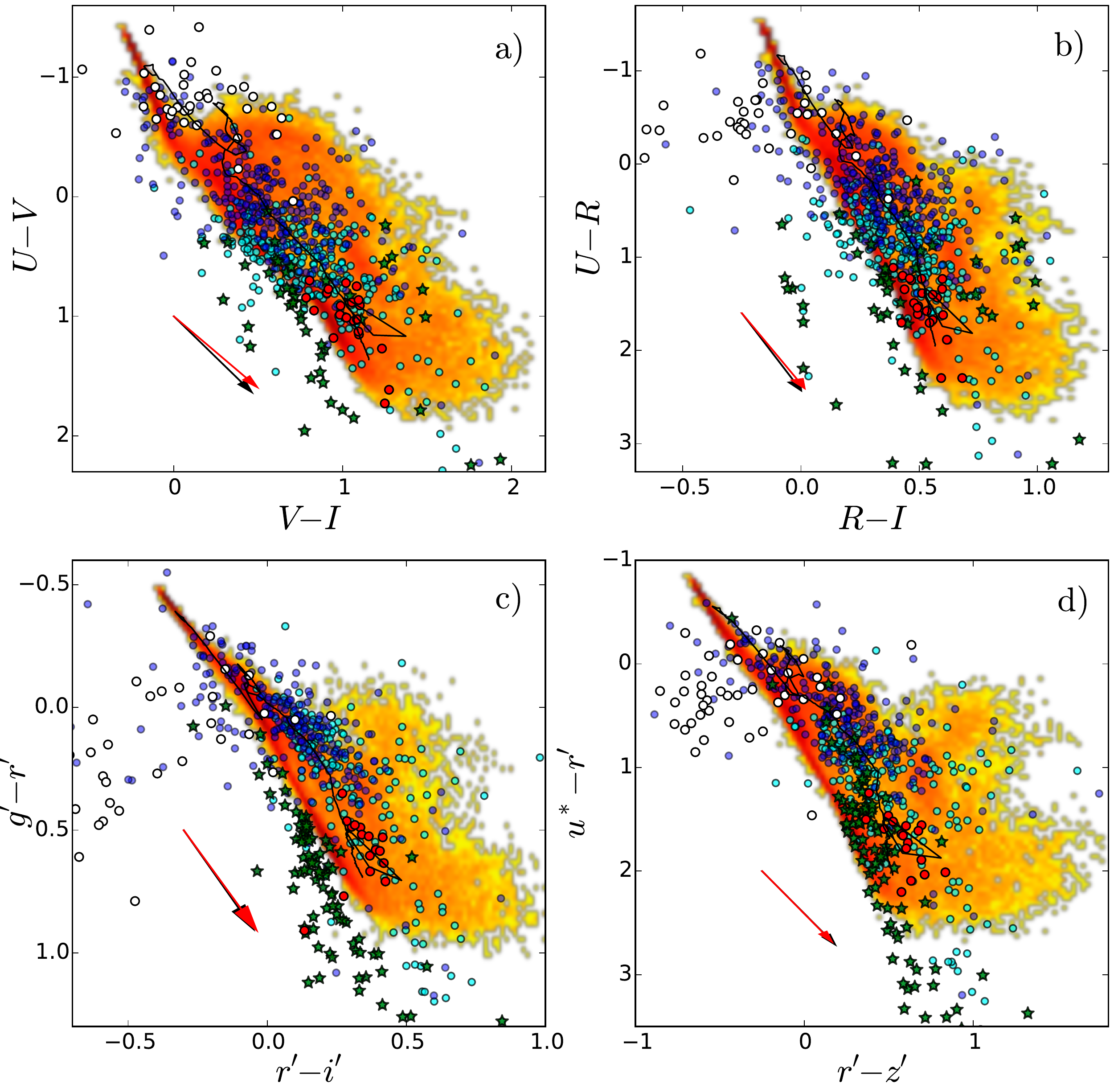}
\caption{\small Color-color diagrams of the cluster sample in optical photometric passbands. The first row of panels shows the color-color diagrams in the standard $UBVRI$ photometric system and the second row in the $u^*g'r'i'z'$ one. In all panels, the blue circles are clusters that are bright in UV, cyan circles are clusters faint in UV, white circles are clusters embedded or close to HII zones, red circles are globular-like clusters, and green stars are likely stars rather than clusters. In each panel, the extent and density of the stochastic star cluster model grid is shown as a density surface (displayed in logarithmic scale), and the solid line is the SSP model for comparison. In both stochastic and SSP models shown here, the metallicity is $\rm{[M/H]}=-0.4$. The mass of the stochastic cluster models shown here is fixed to $\log_{10}(M/M_{\odot})=3.5$, a typical mass of the low-mass clusters studied in this work, to show the extent of their colors. In each panel, the black arrow shows the Milky Way extinction law direction and the red arrow shows the LMC extinction law direction, both computed for $A_V=1$\,mag.}
\label{fig:CCDs_optical_bands}
\end{figure*}

\begin{figure*}
\includegraphics[scale=0.7]{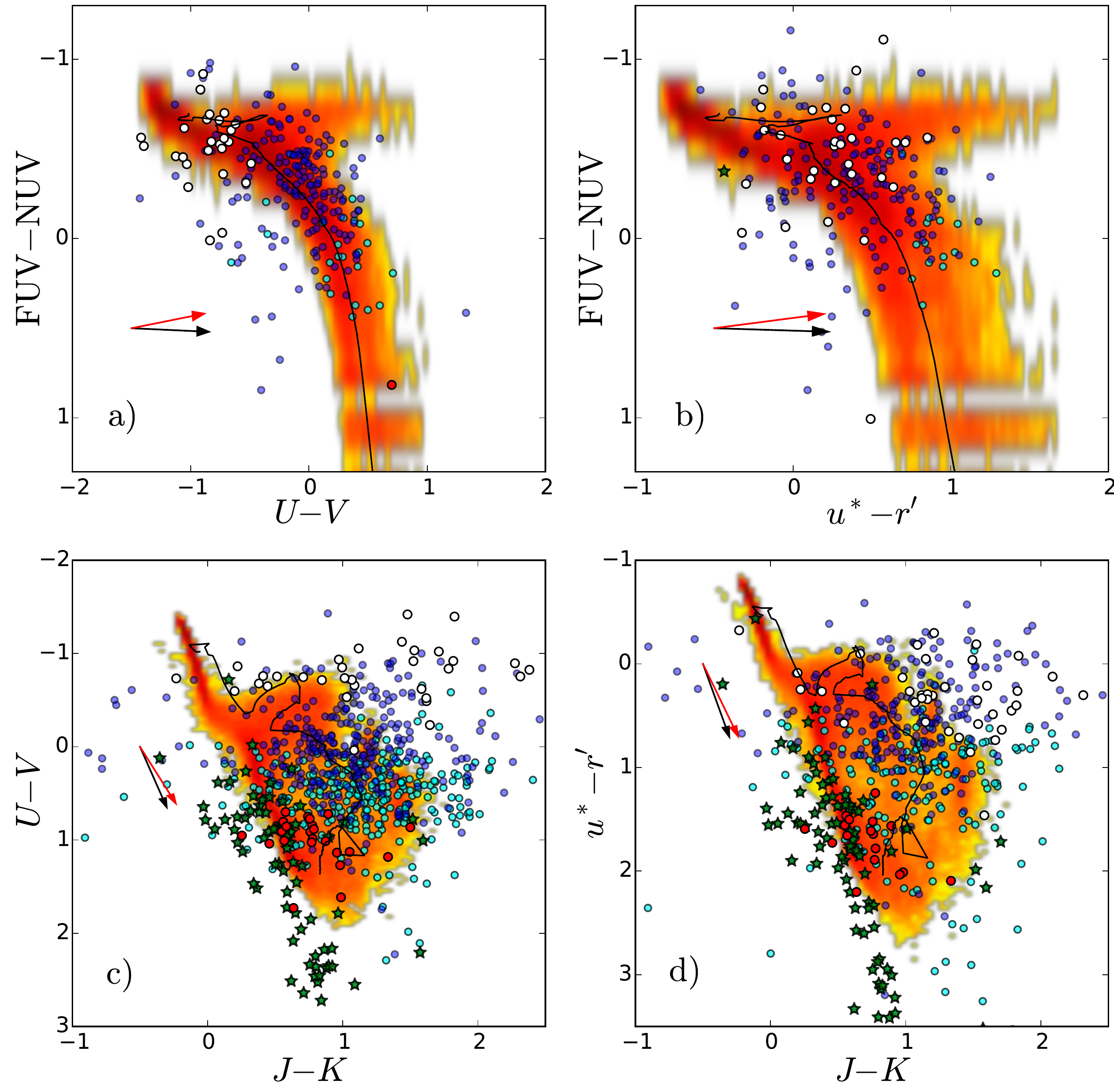}
\caption{\small Same as in Fig. \ref{fig:CCDs_optical_bands}, but for GALEX-optical colors (top panels) and optical-2MASS colors (bottom panels). In the top panels, a small part of the clusters is shown because most of the clusters are too dim in GALEX passbands, or situated too close to bright neighbor objects.}
\label{fig:CCDs_UV_IR_bands}
\end{figure*}

We also add ultraviolet aperture photometry from GALEX\footnote{GALEX: https://archive.stsci.edu/prepds/galex$\_$atlas/index.html} by using aperture radii  of 3\arcsec in both far-ultraviolet (FUV) and near-ultraviolet (NUV) passbands. This photometry was not used to derive the star cluster parameters, but only for the qualitative confirmation of results in case of young clusters, as ultraviolet magnitudes fade very quickly with age, becoming too faint at the distance of M33 after $\gtrsim$100\,Myr. Also, the very wide Point Spread Function (PSF) of the GALEX telescope makes the accurate derivation of UV colors impossible for all clusters, but only for the few relatively isolated ones (at least at a distance of two aperture radii distance from any other UV-emission in the worst cases). 

We also used deep $BVRIH_{\alpha}$ optical images from Subaru 8\,m telescope and 24\,$\mu m$ image from the Spitzer\footnote{Spitzer: http://ssc.spitzer.caltech.edu/spitzerdataarchives/} telescope. We created multi-passband images for each cluster in our catalog that were used to visually confirm the results of our method of star cluster parameter derivation described in Section \ref{sec:parameters_determination}.

Figure\,\ref{fig:CCDs_optical_bands} presents the multi-passband photometric data in different color-color diagrams in optical cases ($UBVRI$ and $u^*g'r'i'z'$). Figure\,\ref{fig:CCDs_UV_IR_bands} shows the clusters in GALEX photometry (top panels) only for objects undisturbed by close neighboring UV emission, and in deep 2MASS 6X photometry (bottom panels). The clusters are shown with SSP models (solid lines) and also with a grid of artificial star cluster models, which take the stochastic dispersion of their colors into account, as described in Section\,\ref{sec:parameters_determination}. Both SSP model and stochastic model grid are shown with the same metallicity, $\rm{[M/H]}=-0.4$. 

Although GALEX photometry is inaccurate for most of the 910 objects because of the presence of possible UV emitting neighboring objects, we used the FUV photometry as a criterion for the qualitative evaluation of the UV emission strength. Objects are said to be bright in UV when their aperture photometry within an aperture radius of 3\arcsec\,is brighter than 20\,mag, and faint in UV when it is fainter than this limit.  

In Figs.\,\ref{fig:CCDs_optical_bands} and \ref{fig:CCDs_UV_IR_bands}, the clusters are color-coded according to their types: bright in UV (blue), faint in UV (cyan), embedded or close to HII zones (white), and globular-like clusters (red). Clusters are classified as globular-like depending on their visual confirmation using HST images \citep{Sarajedini1998,Chandar1999,SanRoman2009}. In addition, the deep Subaru images were used to reject 95 highly probable stars from our star cluster sample. These stellar objects are marked as green star symbols in the color-color diagrams in Figs.\,\ref{fig:CCDs_optical_bands} and \ref{fig:CCDs_UV_IR_bands}. 

After the rejection of these 95 stars from the 910 objects in the sample, as well as a few clusters with incomplete photometry, 747 clusters remain to be studied.

%%%%%%%%%%%%%%%%%%%%%%%%%%%%%%%%%%%%%%%%%%%%%%%%%%%%%%%%%
\section{Method of derivation of cluster parameters}
\label{sec:parameters_determination}
%%%%%%%%%%%%%%%%%%%%%%%%%%%%%%%%%%%%%%%%%%%%%%%%%%%%%%%%%
Following \cite{Fouesneau2010,Fouesneau2014}, and Papers I, II, and III, the derivation of the physical parameters (age, mass, extinction, and metallicity\footnote{Hereafter we refer to extinction and metallicity as $E(B-V)$ and $\rm{[M/H]}$.}) of a given observed star cluster is based on a comparison of its integrated broad-band photometry to a four--dimensional grid (for the age, mass, extinction, and metallicity) of star cluster models. Each node of the grid contains 1\,000 star cluster models. Each star cluster model is built by randomly sampling the stellar mass according to the IMF \citep{Kroupa2001} following the method described in \cite{Deveikis2008} \citep[see also][]{Santos1997,Cervino2002}. The luminosities of clusters were derived using the PADOVA isochrones\footnote{PADOVA isochrones from ``CMD 2.6'': http://stev.oapd.inaf.it/cmd} from \cite{Marigo2008} with the addition of the Thermally Pulsing Asymptotic Giant Branch (TP-AGB) phase from \cite{Girardi2010}.
The grid was built according to the following nodes: from $\log_{10}(t/\mathrm{yr})=6.6$ to 10.1 in steps of 0.05, from $\log_{10}(M/M_{\odot})=2$ to 7 in steps of 0.05, and for 13 metallicities: from $\rm{[M/H]}$ = $+0.2$ to $-2.2$ in steps of 0.2. This results in a grid of 71 values of age, 101 values of mass, with $1\,000$ models per node, hence $\sim$$7\times10^{6}$ models for each metallicity. To limit the number of models that need to be stored in computer memory, the extinction was computed when the observed cluster was compared with the grid of models. It ranges from $E(B-V)=0$ to 1 in steps of 0.01, therefore 101 values for the extinction.
We used the \cite{Gordon2003} extinction law derived for the Large Magellanic Cloud (LMC), as the M33 galaxy is believed to have a similar metallic content and hence may follow a similar extinction law.

We evaluated the likelihood of each node of the grid to represent the magnitudes of a given observed cluster. Within each node, we first computed the likelihood of each of the $1\,000$ star cluster models by
\begin{equation}
L_{\mathrm{model}} = \prod_{f=1}^{F} \frac{1}{ \sqrt{2 \pi}\, \sigma_{f}} \exp \left[ - \frac{\left(\mathrm{mag}_{f,\mathrm{obs}}-\mathrm{mag}_{f,\mathrm{model}}\right)^{2}}{2\,\sigma_{f}^{2}} \right]\,,
\label{eq:L_model}
\end{equation}
where $f$ stands for one particular filter, $\mathrm{mag}_{f}$ for the observed and model magnitudes in that filter, and $F$ for the total number of filters. For example, $F=5$ for the $UBVRI$ photometric system we use in this study. Then the likelihood of the node of age $t$, mass $M$, extinction $E(B-V)$, and metallicity $\rm{[M/H]}$ is the sum of the likelihoods of its models,
\begin{equation}
L_{\mathrm{node}}\left(t,M,E(B-V),\rm{[M/H]}\right) = \sum_{n=1}^{N} L_{\mathrm{model},\,n}\,,
\label{eq:node_likelihood}
\end{equation}
where $N=1\,000$, the total number of models contained in the node. The procedure is repeated for each node of the four--dimensional grid, and the observed star cluster is then classified with the parameters of the node, which maximizes the quantity $L_{\mathrm{node}}$. We note that this procedure could also be applied by using colors (e.g. $U-B$, $u^*-g'$, or other passband combinations) in place of individual magnitudes as the variable $\mathrm{mag}_{f}$ of Eq. \ref{eq:L_model}.

%%%%%%%%%%%%%%%%%%%%%%%%%%%%%%%%%%%%%%%%%%%%%%%%%%%%%%%%%
\section{Artificial tests}
\label{sec:artificial_test}
%%%%%%%%%%%%%%%%%%%%%%%%%%%%%%%%%%%%%%%%%%%%%%%%%%%%%%%%%

\begin{figure*}
\includegraphics[scale=0.485]{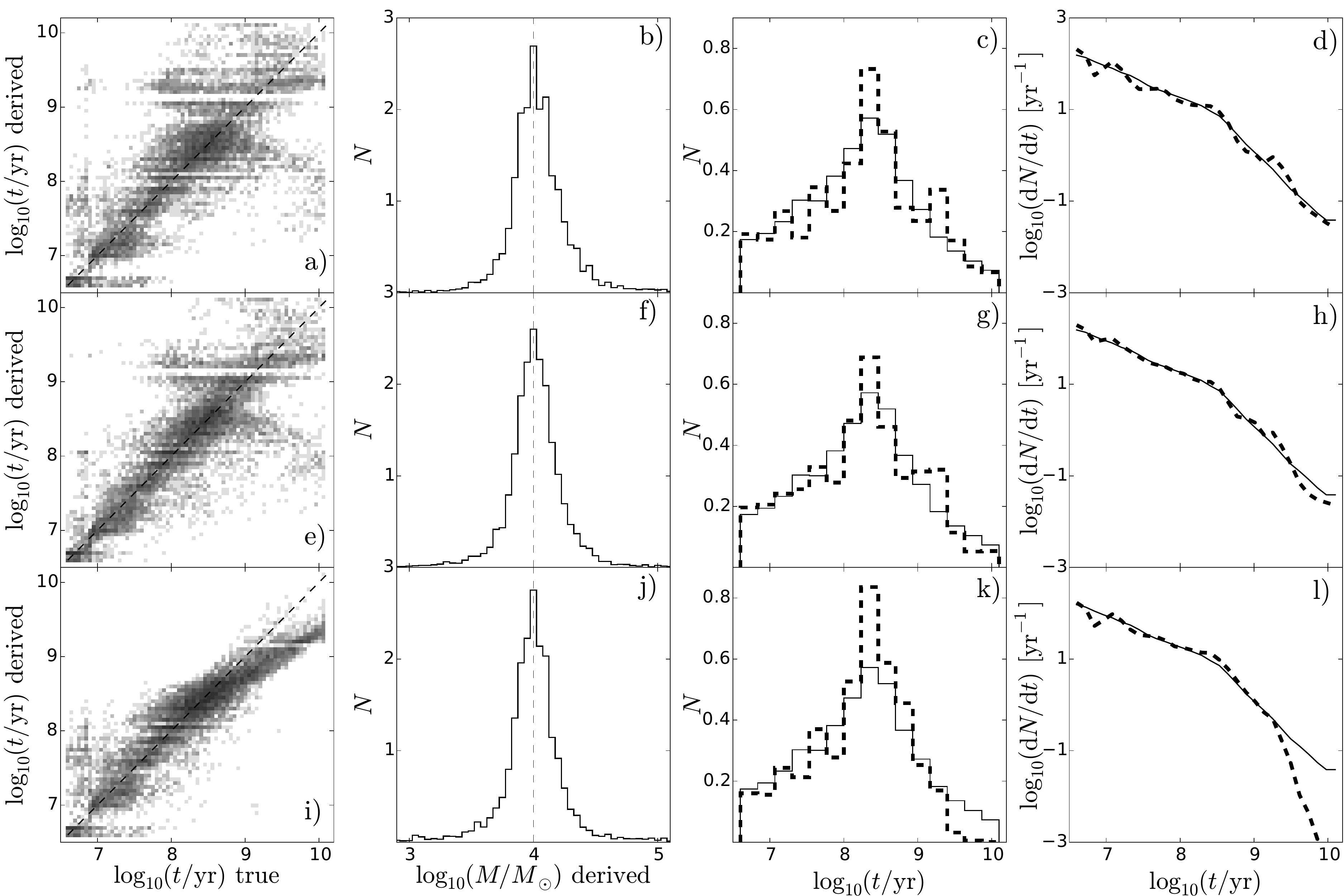}
\caption{\small Artificial tests for $10\,000$ clusters with true age and metallicity following a defined age-metallicity relation (see text), and with a true mass fixed to $\log_{10}(M/M_{\odot})=4$. The first column of panels displays the age derived vs the true age, the second column shows the derived mass distribution, the third column shows the true age distribution (solid line histogram) and derived age distribution (dashed line histogram), and the last column shows the differential true age distribution (solid line), which is composed of a two-slope profile, and the derived differential age distribution (dashed line). The first row of panels shows results obtained using the $UBVRI$ photometric system, the second row shows the results obtained using the $UBVRI\,+\,JHK$ system and the last row shows the results obtained with the GALEX\,+\,$UBVRI$ system. Here the metallicity of the model grid is fixed to $\rm{[M/H]}=-0.4$.}
\label{fig:AT_Zm04}
\end{figure*}

\begin{figure*}
\includegraphics[scale=0.485]{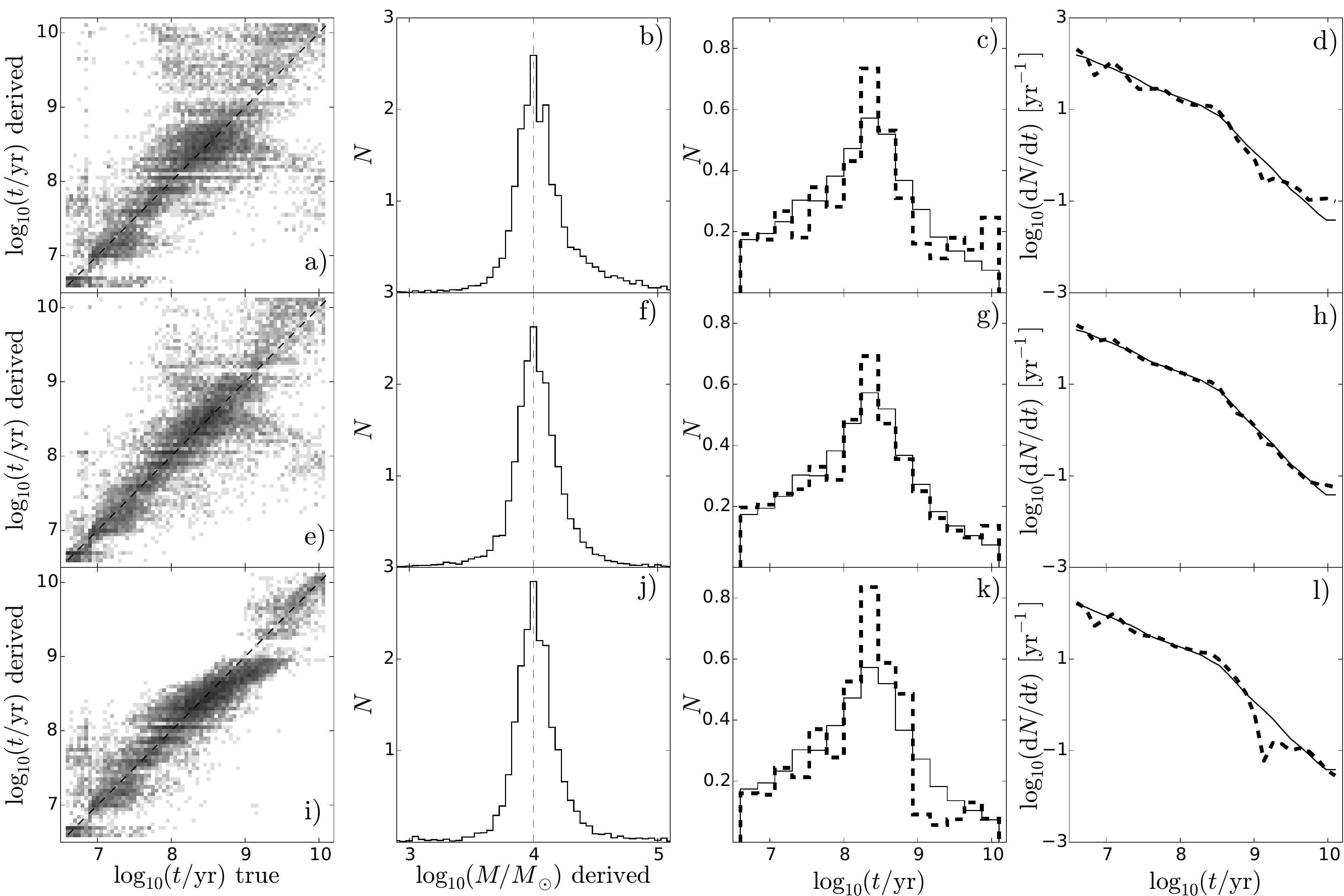}
\caption{\small Same as in Fig. \ref{fig:AT_Zm04}, but here, for clusters which have derived ages larger than $\log_{10}(t/\mathrm{yr})=9$ when using fixed $\rm{[M/H]}=-0.4$ metallicity, we re-derive a solution leaving the metallicity free to vary in the range $[+0.2,-2.2]$.}
\label{fig:AT_Zm04_Zfree_after1Gyr}
\end{figure*}

\begin{figure*}
\includegraphics[scale=0.485]{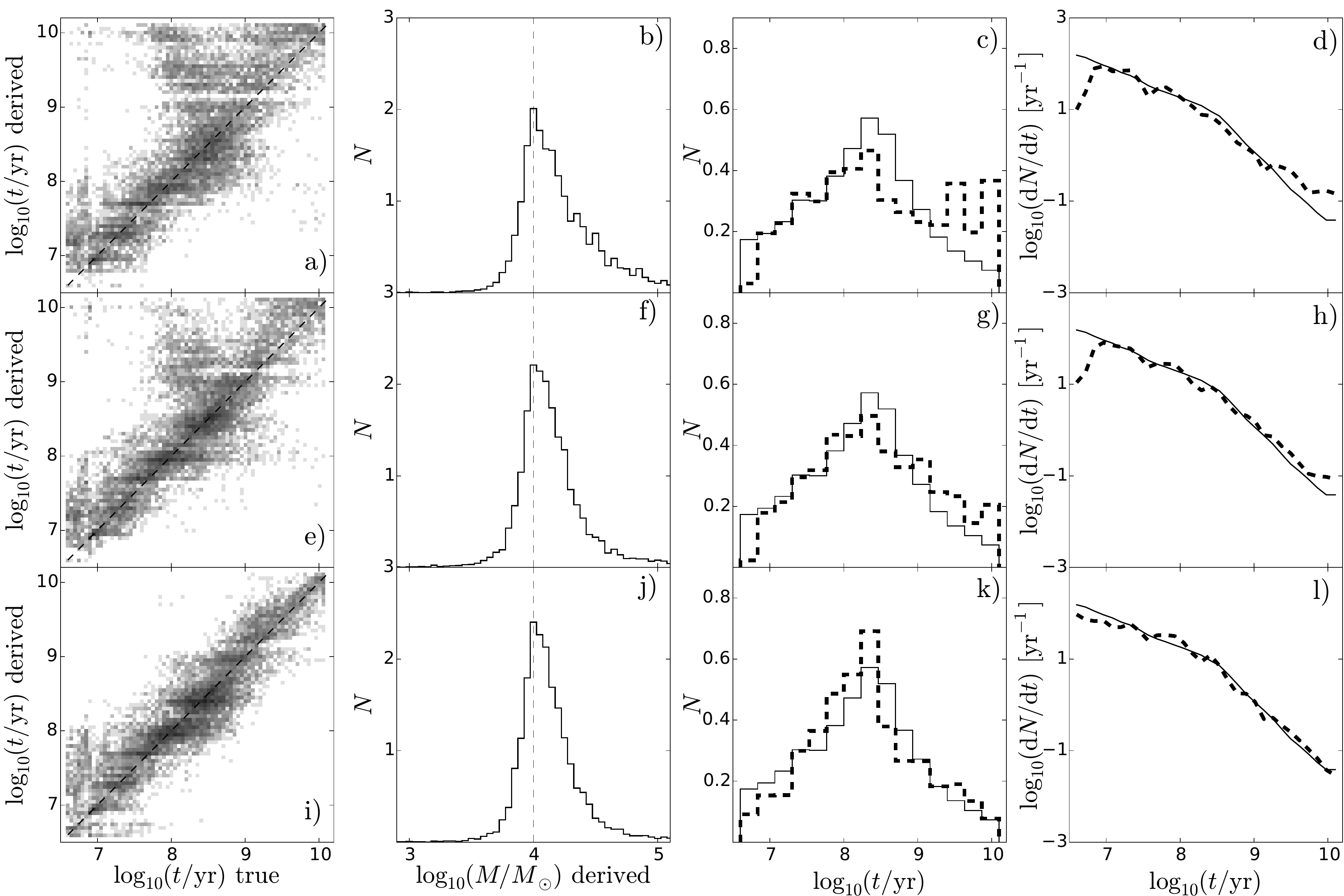}
\caption{\small Same as in Fig. \ref{fig:AT_Zm04}, but here the metallicity of the model grid is left free over the whole age range.}
\label{fig:AT_Zfree}
\end{figure*}

The ability of the method to derive star cluster parameters has already been evaluated in Papers I, II, and III. Here we are interested in seeing for which conditions the method would be sensitive enough to detect a change of slope in the number of clusters per age bin distribution (hereafter referred as \emph{differential age distribution}) of the cluster sample such as shown by the solid line in Fig.\,\ref{fig:AT_Zm04}d. Indeed, a two-slope profile in the differential age distribution could be interpreted as a decrease in the number of clusters due to an evolutionary fading of the cluster magnitudes (first slope), and a decrease of the number of clusters due to their disruption (second slope), as is discussed in greater details in Sect.\,\ref{sec:derived_parameters}. Here the objective is to model an artificial star cluster population with such a two-slope profile in the true differential age distribution and to see whether the derived differential age distribution reproduces this profile depending on which photometric system we use.

We generated a sample of $10\,000$ artificial star clusters. The differential age distribution of the artificial clusters was chosen to mimic a two-slope profile similar to that described in \cite{Vansevicius2009} for M31 star clusters (see their Fig. 5a, reproduced in our Fig.\,\ref{fig:AT_Zm04}d in solid line). For simplicity, the mass of the clusters was fixed to $\log_{10}(M/M_{\odot})=4$, a typical value for the clusters observed in M33. The extinction was randomly generated uniformly in the range $E(B-V)=0$ to 1. 

For the cluster metallicities, we use a very simple age-metallicity relation: for the youngest clusters the metallicity is supersolar, $\rm{[M/H]}=0.2$, and for the oldest clusters the metallicity is very low, $\rm{[M/H]}=-1.8$. The age-metallicity relation is linear between these values in the age ($\log_{10}(t/\mathrm{yr})$)--metallicity ($\rm{[M/H]}$) space. The metallicity of clusters is generated with a Gaussian dispersion of 0.4\,dex standard deviation around this age-metallicity relation.  

The first test, presented in Fig.\,\ref{fig:AT_Zm04}, was performed using a cluster model grid with the metallicity fixed to $\rm{[M/H]}=-0.4$ to show the possible biases that occur when metallicity is not a free parameter. The test was performed for three photometric system combinations: optical ($UBVRI$), optical with near-infrared ($UBVRI\,+\,JHK$) and ultraviolet with optical (GALEX$\,+\,UBVRI$). The $UBVRI$ system was used as a reference for the mass derivation, so it was used in magnitudes in the Eq.\,\ref{eq:L_model}. For the other passbands used we have used colors instead, FUV-NUV for the GALEX passbands and $J-H$, $J-K$, and $H-K$ for 2MASS. This allowed us to combine different catalogs of clusters built with slightly different aperture sizes: we used the magnitudes for one catalog, and the colors for the others. However it is important that for each cluster at least one magnitude should be given, not just colors, so that the mass of the clusters could be estimated reliably by the method.

We added photometric uncertainties as a Gaussian noise with standard deviations of 0.05\,mag for each $UBVRI$ photometric passbands, 0.1\,mag for $JHK$, and 0.15 mag for GALEX FUV and NUV. 

In Fig.\,\ref{fig:AT_Zm04} we see how the derived cluster age (panel a) and mass (panel b) are distributed versus the true values (indicated by dashed lines). Fig.\,\ref{fig:AT_Zm04}c and Fig.\,\ref{fig:AT_Zm04}d concentrate on the age derivation, and show that the peak in the true age distribution (indicated by thin lines in panels c and d) has already been found using optical data only. The true peak in age, situated at $\log_{10}(t/\mathrm{yr})\sim8.3$ in Fig.\,\ref{fig:AT_Zm04}c (solid line histogram) and corresponding to a change of slope at $\log_{10}(t/\mathrm{yr})=8.5$ in Fig.\,\ref{fig:AT_Zm04}d (solid line) is correctly derived as maximum in Fig.\,\ref{fig:AT_Zm04}c (dashed line histogram) and change of slope in Fig.\,\ref{fig:AT_Zm04}d (dashed line). However, the apparent good match between true age and derived age distributions in Fig.\,\ref{fig:AT_Zm04}c and Fig.\,\ref{fig:AT_Zm04}d can be rather misleading when using $UBVRI$ photometry alone. Indeed the direct comparison of the individual clusters' true and derived age in Fig.\,\ref{fig:AT_Zm04}a shows that the agreement is far from evident, especially at old ages, where the true metallicity of artificial clusters and the fixed value of the model grid ($\rm{[M/H]}=-0.4$) deviate most. As a natural consequence, most of the clusters with true age above $\log_{10}(t/\mathrm{yr})=9.5$ have age underestimated. Also, the presence of the natural age-extinction degeneracy in the optical $UBVRI$ case, already discussed in papers I and II, produces the streaks developing perpendicularly to the left and to the right of the diagonal identity dashed line in Fig.\,\ref{fig:AT_Zm04}a. The situation is less extreme, but still strongly affected by these degeneracies, when we add $JHK$ passbands to $UBVRI$ ones (second row of panels). When we use GALEX with $UBVRI$ passbands (third row of panels), the age-extinction degeneracy disappears, but the deviation from the identity line occurs because of the strong sensitivity of ultraviolet to the metallicity. \cite{Bianchi2011} indeed shows by use of integrated spectra of simple stellar population models that it is in the ultraviolet spectral region that the spectra are most affected by a change in metallicity. 

We performed a second test, fixing the metallicity to $\rm{[M/H]}=-0.4$ for all clusters, and then, only for clusters which have derived age larger than $\log_{10}(t/\mathrm{yr})=9$, we re-derive a solution leaving the metallicity free to vary in the range $[+0.2,-2.2]$. Indeed one notices in the first test that the situation was most complicated for clusters with true age above 1 Gyr. The results, shown in Fig.\,\ref{fig:AT_Zm04_Zfree_after1Gyr} still suffer from strong age-extinction degeneracy in the case of $UBVRI$ passbands only (first row of panels). The inclusion of near-infrared photometry improves much the derivation as the streaks developing perpendicularly to the identity line in Fig.\,\ref{fig:AT_Zm04_Zfree_after1Gyr}e are strongly reduced. In this case, the match between the true and derived age distributions in Fig.\,\ref{fig:AT_Zm04_Zfree_after1Gyr}g and Fig.\,\ref{fig:AT_Zm04_Zfree_after1Gyr}h is much more secure for all age ranges. When using GALEX with $UBVRI$ (third row of panels), a gap is visible in Fig.\,\ref{fig:AT_Zm04_Zfree_after1Gyr}i due to the strong sensitivity of the ultraviolet flux to metallicity. In this case, strong biases are still present in the age distribution (Fig.\,\ref{fig:AT_Zm04_Zfree_after1Gyr}k and Fig.\,\ref{fig:AT_Zm04_Zfree_after1Gyr}l).

A third test was performed in which metallicity of the model grid was left free in the whole age range, and the results are presented in Fig.\,\ref{fig:AT_Zfree}. Here we see that the use of optical passbands only (first row of panels) or even optical with near-infrared (second row) can lead to strong biases as these photometric systems are not sensitive enough to discriminate between models of different metallicities (see also Papers II and III for the sensitivity of the derived parameters on the metallicity, as well as for the derivation of the metallicity parameter). As a consequence, age distributions (panels c and d for $UBVRI$ case, panels g and h for $UBVRI\,+\,JHK$ case) are strongly affected. Only ultraviolet associated with optical data passbands are able to break the age-extinction degeneracies when metallicity is left free, as shown in last row of panels. As a consequence, the derivation of the correct two-slope profile in the differential age distribution is best done using GALEX\,+\,$UBVRI$ when metallicity is left free, see Fig.\,\ref{fig:AT_Zfree}l).

%%%%%%%%%%%%%%%%%%%%%%%%%%%%%%%%%%%%%%%%%%%%%%%%%%%%%%%%%
\section{Derived physical parameters of clusters}
\label{sec:derived_parameters}
%%%%%%%%%%%%%%%%%%%%%%%%%%%%%%%%%%%%%%%%%%%%%%%%%%%%%%%%%
We applied the method of derivation of physical parameters to the sample of 747 M33 clusters using the optical $UBVRI$ and near-infrared $JHK$ passbands. We first fixed the metallicity to $\rm{[M/H]}=-0.4$ for all clusters, and then, only for clusters that have derived ages larger than $\log_{10}(t/\mathrm{yr})=9$, we re-derive a solution leaving the metallicity free to vary in the range $[+0.2,-2.2]$, as was shown to be the best choice for this passband combination in the previous section. 

As was done for the artificial tests, we used the $UBVRI$ system as a reference for the mass derivation, and so it was used in magnitudes in Eq.\,\ref{eq:L_model}. For the other passbands used, $u^*g'r'i'z'$ and $JHK$, we used colors instead to avoid problems of different apertures in the different catalogs used. The colors used were $u^*-g'$, $g'-r'$, $g'-i'$, $r'-i'$, and $i'-z'$ for the CFHT passbands, and $J-H$, $J-K$, and $H-K$ for the 2MASS passbands. 

We used the extinction law of \cite{Gordon2003} derived for the LMC, assuming that for a similar metallic content the M33 galaxy would have a similar extinction law. The minimum extinction of clusters was set to $E(B-V)=0.04$ mag, the value of the foreground extinction in the direction of M33 estimated from the \cite{Schlegel1998} extinction maps. 

\begin{figure*}
\includegraphics[scale=0.362]{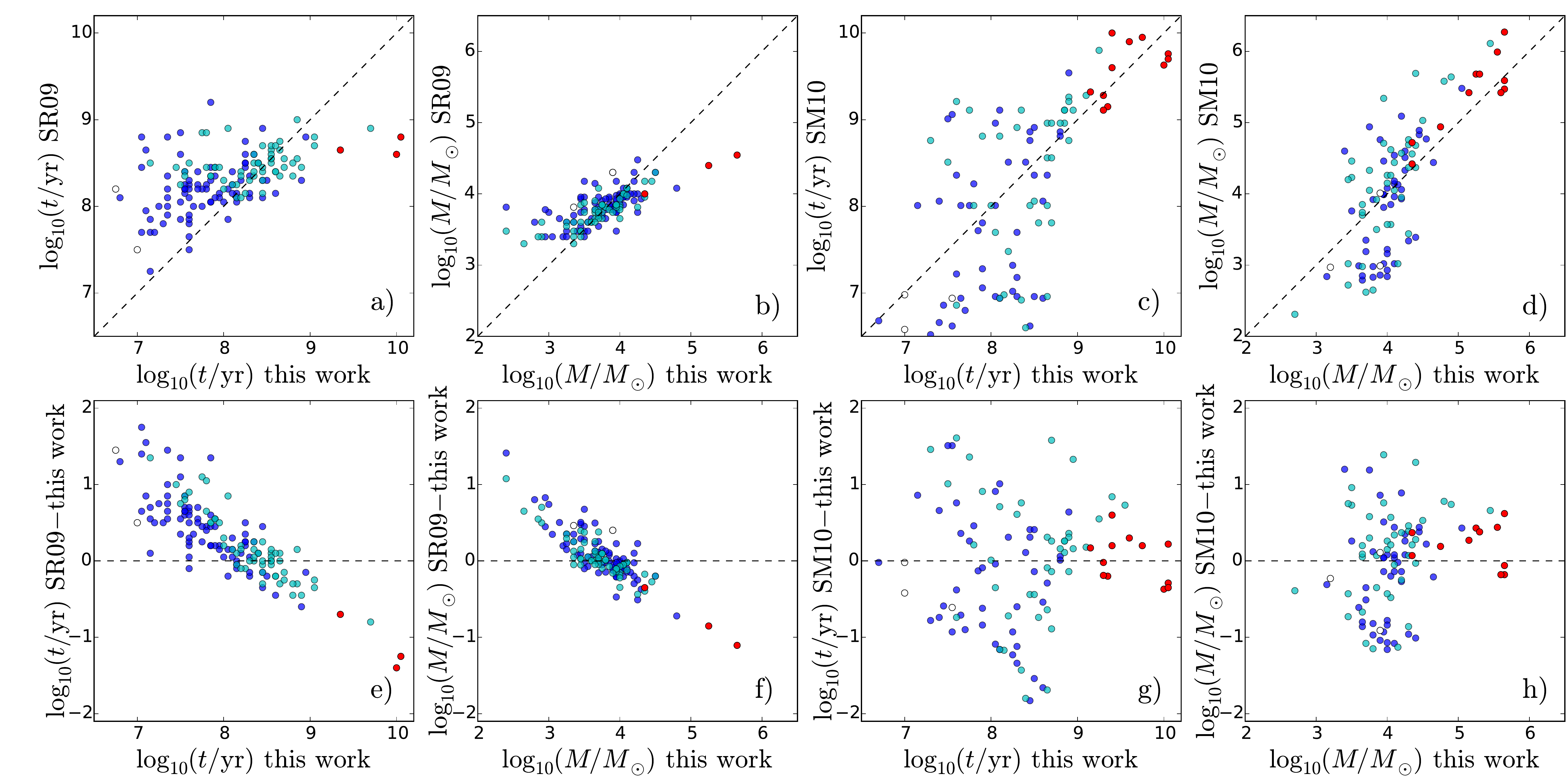}
\caption{\small Comparison of the age and mass derived for clusters common with \cite{SanRoman2009} (panels \textbf{a},\textbf{b},\textbf{e},\textbf{f}), and for clusters common with \cite{Sarajedini2007} (panels \textbf{c},\textbf{d},\textbf{g},\textbf{h}). The first row shows the comparison of the parameters of these studies vs this work, and the second row shows the difference of the parameters derived in their studies and this work vs the parameters derived in this work. The color-coding of clusters is the same as defined in Fig.\,\ref{fig:CCDs_optical_bands}.}
\label{fig:Comparison_parameters}
\end{figure*}

The results obtained here are compared in Fig.\,\ref{fig:Comparison_parameters}a,b,e,f for the age and mass with the ones of 160 clusters of \cite{SanRoman2009}, obtained by isochrone fitting on HST-resolved color-magnitude diagrams. In the case of the \cite{SanRoman2009}, cluster ages are enclosed in a much narrower range, mainly between 50\,Myr and 1\,Gyr. Although our age distribution is wider than in their case, a satisfactory agreement is found between both sets of results, as well as for the mass parameter. 

Globular-like clusters (red circles, visually confirmed as globular clusters on HST images) are found to be very old in our case, and more massive. \cite{SanRoman2009} noted that the lack of clusters with ages older than $\sim$1\,Gyr in their catalog is linked to the fact that their resolved color-magnitude diagrams are generally not deep enough to detect the main sequence turn-off of clusters older than this age. In Fig.\,\ref{fig:Comparison_parameters}a,b,e,f, the two oldest and most massive clusters (according to our derived age and mass) are globular clusters known as CBF85-U137 and MKKSS12. For CBF85-U137, we find $\log_{10}(t/\mathrm{yr})=10.05$ and $\log_{10}(M/M_{\odot})=5.25$ while \cite{SanRoman2009} give $\log_{10}(t/\mathrm{yr})=8.8$ and $\log_{10}(M/M_{\odot})=4.4$. \cite{Chandar2002} give $\log_{10}(t/\mathrm{yr})=10.08$ (12\,Gyr in their table 5) using spectroscopic line index models of \cite{Worthey1994}. For MKKSS12, we find $\log_{10}(t/\mathrm{yr})=10.0$ and $\log_{10}(M/M_{\odot})=5.65$ while \cite{SanRoman2009} give $\log_{10}(t/\mathrm{yr})=8.6$ and $\log_{10}(M/M_{\odot})=4.54$. \cite{Chandar2002} give $\log_{10}(t/\mathrm{yr})=9.4$ using SSP models, and \cite{Ma2004} give $\log_{10}(t/\mathrm{yr})=9.63$ and $\log_{10}(M/M_{\odot})=5.47$ using their BATC spectrophotometric system composed of 13 narrow passbands, also compared to SSP models.

For young clusters, the age given by \cite{SanRoman2009} is often older than in our values. In Fig.\,\ref{fig:Comparison_parameters}a,e we note that the two white circles, which are clusters still within HII zones, and so very young, are also seen as older in \cite{SanRoman2009} than in our study. Also, many clusters that are bright in UV (blue circles) are also older in \cite{SanRoman2009} than in our case. However, UV brightness fades very quickly, becoming faint at the distance of M33 after $\gtrsim$100\,Myr, making it unlikely that the age of these clusters is older.

Fig.\,\ref{fig:Comparison_parameters}c,d,g,h compares the results found for clusters common with the \cite{Sarajedini2007} merged catalog (labeled ``SM10'' in the figure because revised in 2010); the agreement for both age and mass parameters is not as good. One has to keep in mind that the \cite{Sarajedini2007} catalog contains results from very different studies. Many of theses results have been obtained using SSP models on integrated unresolved ground-based photometry using optical colors only, a technique that has been shown to be strongly biased by the stochasticity problem and the presence of strong degeneracies \citep[see, e.g.,][Papers I and II]{Fouesneau2010}. \cite{Fouesneau2010} created a sample of stochastically generated artificial star clusters and tried to derive their age using SSP models. Their results, shown in the bottom left panel of their Fig. 3, are similar to the values given in our Fig.\,\ref{fig:Comparison_parameters}c. The concentrations of solution at ages $\log_{10}(t/\mathrm{yr})\sim7$ and $\log_{10}(t/\mathrm{yr})\sim9$ for \cite{Sarajedini2007} seem to be artifacts due to the SSP method, as is the case for the SSP derived ages in Fig. 3 of \cite{Fouesneau2010}. \cite{Popescu2012} show the same features, this time for real clusters. They derived the age of LMC star clusters using the stochastic method and compared these values to the ages derived by \cite{Hunter2003} using the SSP method. The comparison, shown in Fig. 8 of \cite{Popescu2012} (the order of the x-- and y--axes is flipped compared to our figure) shows similar features to the \cite{Fouesneau2010} study and to this work: the deviations from the identity line are attributed to artifacts of the SSP method of parameter derivation.

\begin{figure*}
\includegraphics[scale=0.455]{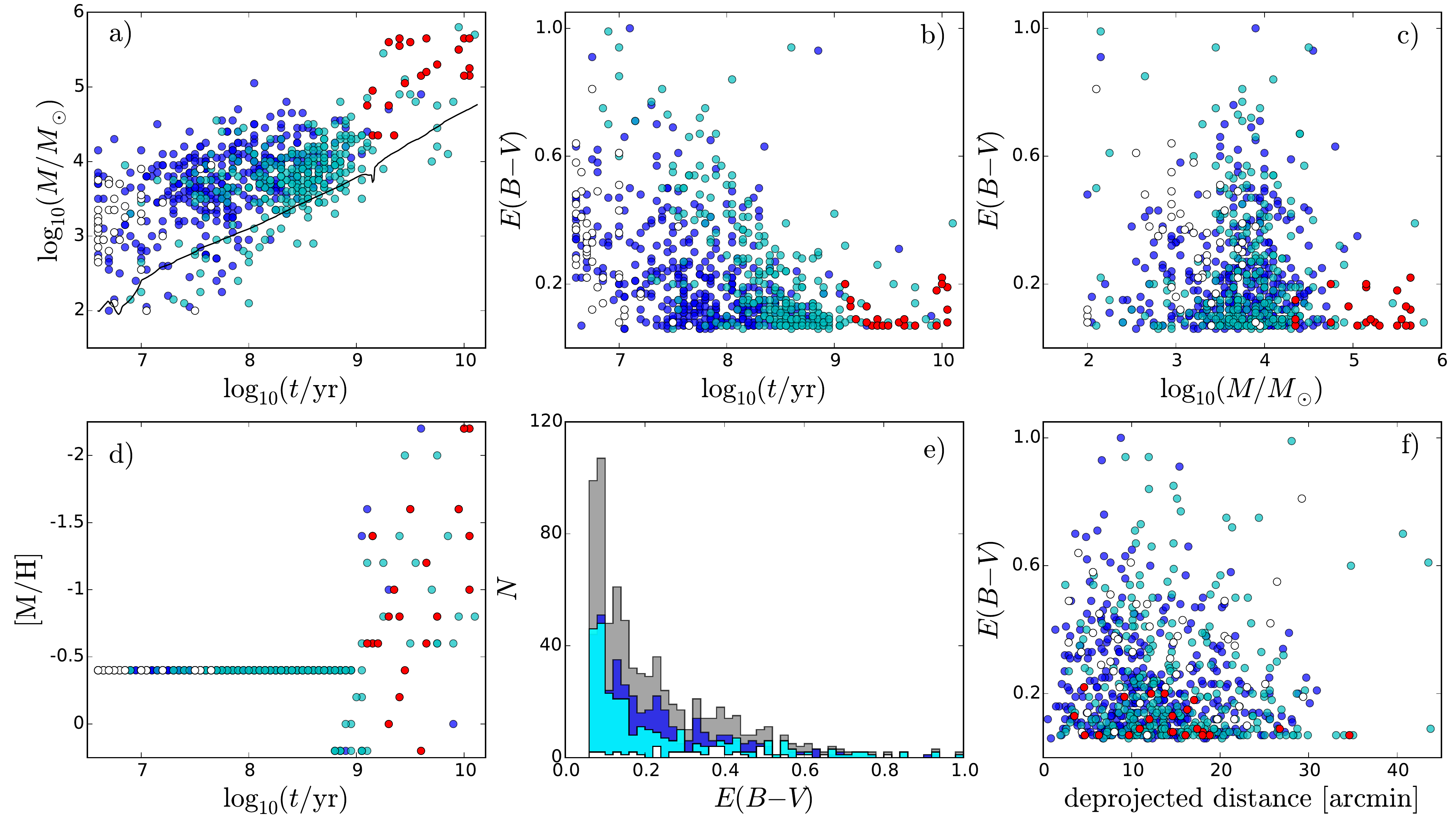}
\caption{\small The first row of panels shows the mass vs age (panel \textbf{a}), extinction vs age (panel \textbf{b}), and extinction vs mass (panel \textbf{c}) derived for the 747 clusters studied in this paper. The solid line in panel \textbf{a} shows the photometric 50\% completeness limit estimated in \cite{Fan2014} for their optical ground-based cluster catalog. The second row shows the metallicity vs the age (panel \textbf{d}), the extinction histogram (panel \textbf{e}) and the extinction vs the deprojected galactocentric distance (panel \textbf{f}) of the cluster population. As emphasized in the text, the metallicity of the clusters, shown in panel \textbf{d}, has been fixed to $\rm{[M/H]}=-0.4$ when the derived ages using that metallicity is lower than 1\,Gyr, and for clusters with older derived ages, a new solution is derived with free metallicity in the range $[+0.2,-2.2]$. The color-coding of clusters is the same as defined in Fig.\,\ref{fig:CCDs_optical_bands}.}
\label{fig:Parameters_global}
\end{figure*}

Figure\,\ref{fig:Parameters_global} presents the results for the star cluster sample studied in the paper. As expected, the mass versus age distribution in Fig.\,\ref{fig:Parameters_global}a shows a different typical age for the different classes of clusters. The youngest ones are the embedded or close to HII zones (white circles), then come the clusters bright in the UV (blue circles), then the faint in the UV (cyan circles), and finally the globular-like clusters (red circles). The solid line shows the 50\% completeness line evaluated in \cite{Fan2014} for their cluster sample. As our merged cluster sample also contains HST detected objects from \cite{SanRoman2009}, some clusters may be found well below this limit.
 
For M33, \cite{U2009} derived the extinction of 22 supergiant stars, which resulted in an extinction distribution centered on $E(B-V)$ = 0.1\,mag. \cite{U2009} also used the data of \cite{Rosolowsky2008} to derive $E(B-V)$ values for 58 HII regions, and show in their Fig.\,9 that extinction can be expected to be $E(B-V)$ $\lesssim$ 0.3\,mag for those regions (except for three objects), with an average $E(B-V)$ $\sim$ 0.11\,mag.

The extinction of all clusters depending on their deprojected galactocentric distance (assuming that all clusters are in the disk, which is incorrect for globular-like clusters) is shown in Fig.\,\ref{fig:Parameters_global}f. The global median extinction is 0.16\,mag. The extinction that we found is generally higher for young clusters than for old ones, as shown in Fig.\,\ref{fig:Parameters_global}b. Indeed, the majority of clusters still embedded or close to HII zones (white circles) are found to be more extincted with a median of 0.34\,mag. Clusters bright in UV (blue circles) have a median extinction of 0.17\,mag while clusters faint in UV (cyan circles) have a lower median extinction of 0.14\,mag. Globular-like clusters (red circles) have the smallest median extinction with 0.09\,mag.

\begin{figure}
\centering
\includegraphics[scale=0.24]{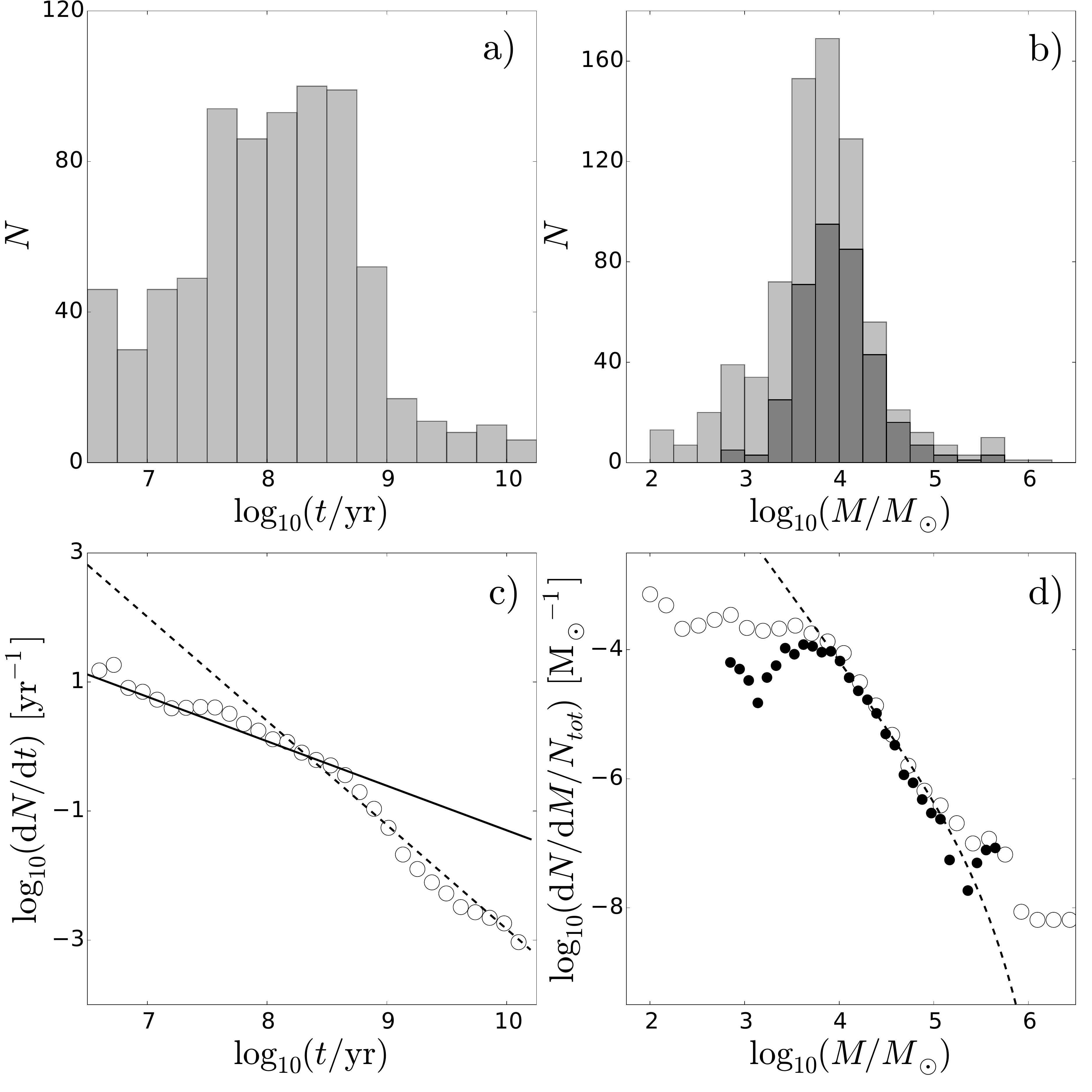}
\caption{\small Top row: age (panel \textbf{a}) and mass (panel \textbf{b}) distributions derived for the M33 star cluster sample. Bottom row: differential age (panel \textbf{c}) and mass (panel \textbf{d}) distributions. In panel \textbf{c}, the solid line and the dashed line are respectively the cluster evolutionary fading rate and the cluster disruption rate, both taken from the case of M31 \citep{Vansevicius2009} and shifted vertically here, for comparison. The dark histogram in panel \textbf{b} and the black circles in panel \textbf{d} represent a subsample of clusters with ages between 100\,Myr and 3\,Gyr. The dashed line in panel \textbf{d} represents the cluster mass function that follows a \cite{Schechter1976} function with $\beta=2$ and $M^{*}= 2\times10^{5} M_{\odot}$.}
\label{fig:Age_Mass_histo-and_Schechter_distributions}
\end{figure}

Figure\,\ref{fig:Age_Mass_histo-and_Schechter_distributions} describes the age and mass distributions (panels a and b) as well as the differential age and mass distributions (panels c and d). We see that the differential age distribution is composed of a two-slope profile. \cite{Boutloukos2003} and \cite{Lamers2005} interpreted the first slope as a natural magnitude fading as a result of stellar evolution, and the second slope as being due to cluster disruption mechanisms such as the galaxy tidal field effect or encounters with giant molecular clouds. Hence we see here that the cluster sample is dominated by the magnitude fading until $\log_{10}(t/\mathrm{yr})\sim 8.5$ and that after the cluster disruption phase takes over. This typical lifetime scale is comparable to that derived for the star cluster population in the southwest field of the M31 galaxy by \cite{Vansevicius2009} using a star cluster sample photometry from \cite{Narbutis2008}. 

Following \cite{Gieles2009}, we compare the cluster differential mass distribution to the \cite{Schechter1976} distribution function, defined as
\begin{equation}
dN/dM = A \times M^{-\beta} \times \exp (-M/M^{*})
\label{eq:Schechter_MF}
\end{equation} 
where $A$ constant scales with the cluster formation rate, $\beta$ is the power-law index of the mass function and $M^{*}$ stands for the characteristic mass after which the exponent term decreases strongly. 

As was found for most spiral galaxies \citep[see, e.g.,][]{Larsen2009,pzwart10}, the derived mass function of the star cluster sample follows the Eq.\,\ref{eq:Schechter_MF} distribution function with $\beta=2$ and $M^{*}= 2\times10^{5} M_{\odot}$. We adapted the scaling constant $A$ to scale it to the cluster mass distribution, shown in Fig. \ref{fig:Age_Mass_histo-and_Schechter_distributions}d.

%%%%%%%%%%%%%%%%%%%%%%%%%%%%%%%%%%%%%%%%%%%%%%%%%%%%%%%%%
\section{Conclusions} 
\label{sec:conclusions}
%%%%%%%%%%%%%%%%%%%%%%%%%%%%%%%%%%%%%%%%%%%%%%%%%%%%%%%%%
We studied the star cluster system of the M33 galaxy, using the most recent optical broad-band photometry catalogs, and supplemented near-infrared measurements using deep 2MASS images. As most of clusters are partially resolved or unresolved, we used a method of star cluster parameter derivation which takes into account the natural dispersion of the integrated colors due to the stochastic sampling of stars in the clusters. We present the derivation of the age, mass, and extinction of the clusters for a metallicity fixed to $\rm{[M/H]}=-0.4$ (LMC-like), and when the age derived is larger than 1\,Gyr, then a new solution is derived using free metallicity in the range $[+0.2,-2.2]$. 

We ensured, by use of artificial clusters, that the star cluster physical parameter derivation method can correctly derive a given two-slope profile in the differential age distribution, testing it for different photometric systems: optical alone ($UBVRI$), optical with near-infrared ($UBVRI\,+\,JHK$), and ultraviolet with optical (GALEX\,+\,$UBVRI$). We showed that the optical with near-infrared case is fit for the correct derivation of the two-slope profile, and we used it for the M33 star cluster system.

A two-slope profile of differential age distribution shows that the typical lifetime before disruption of star clusters in the M33 star cluster system is found to be $\sim 300$\,Myr, comparable to what is found for M31 star clusters. We show that the differential mass distribution of clusters is consistent with a \cite{Schechter1976} function with a power-law index $\beta=2$ and a characteristic mass $M^{*}= 2\times10^{5} M_{\odot}$.

\begin{acknowledgements}
We are grateful to the anonymous referee who helped improve the paper, as well as to T. H. Jarrett (IPAC/Caltech) for making the deep 2MASS 6X $JHK$ images available to us. We also acknowledge the intense use of the Topcat software\footnote{http://www.starlink.ac.uk/topcat/}. This research was funded by a grant (No. MIP-074/2013) from the Research Council of Lithuania.
\end{acknowledgements}

%\bibliographystyle{aa}
%\bibliography{Thesis}

\end{document}